\documentclass[prb,twocolumn,superscriptaddress]{revtex4-2}
\usepackage{amsfonts}
\usepackage{amssymb}
\usepackage{amsmath}
\usepackage{graphicx}
\usepackage{dcolumn}
\usepackage{bm}
\usepackage{units}
\usepackage{multirow}
\usepackage{CJKutf8}
\usepackage{subfigure}
\usepackage{color}%
\usepackage{url}
\usepackage[colorlinks,linkcolor=blue,anchorcolor=blue,citecolor=blue,urlcolor=blue]{hyperref}
\usepackage{array}
\begin{document}
\title{Dancing Synchronization in Coupled Spin-Torque Nano-Oscillators}
\author{H. T. Wu}
\affiliation{Center for Spintronics and Quantum Systems, State Key Laboratory for Mechanical Behavior 
of Materials, Xi'an Jiaotong University, No.28 Xianning West Road, Xi'an, Shaanxi, 710049, China} 
\affiliation{Department of Physics, The Hong Kong University of Science and Technology, 
Clear Water Bay, Kowloon, Hong Kong, China}
	
\author{Lei Wang (\begin{CJK}{UTF8}{gbsn}王蕾\end{CJK})}
\affiliation{Center for Spintronics and Quantum Systems, State Key Laboratory for Mechanical Behavior 
of Materials, Xi'an Jiaotong University, No.28 Xianning West Road, Xi'an, Shaanxi, 710049, China}
	
\author{Tai Min}
\email{tai.min@xjtu.edu.cn}
\affiliation{Center for Spintronics and Quantum Systems, State Key Laboratory for Mechanical Behavior 
of Materials, Xi'an Jiaotong University, No.28 Xianning West Road, Xi'an, Shaanxi, 710049, China} 
	
\author{X. R. Wang}
\email{phxwan@ust.hk}
\affiliation{Department of Physics, The Hong Kong University of Science and Technology, 
Clear Water Bay, Kowloon, Hong Kong, China}
\affiliation{HKUST Shenzhen Research Institute, Shenzhen, 518057, China}
	
\begin{abstract} 
We are reporting a new type of synchronization, termed {\it dancing synchronization}, 
between two spin-torque nano-oscillators (STNOs) coupled through spin waves. 
Different from the known synchronizations in which two STNOs are locked with various fixed 
relative phases, in this new synchronized state two STNOs have the same frequency, but their  
relative phase varies periodically within the common period, resulting in a dynamic waving pattern. 
The amplitude of the oscillating relative phase depends on the coupling strength of two STNOs,  
as well as the driven currents. The dancing synchronization turns out to be universal, and 
can exist in two nonlinear Van der Pol oscillators coupled both reactively and dissipativly. 
Our findings open doors for new functional STNO-based devices.
\end{abstract}
\maketitle
\section{Introduction}
Synchronization is the coordination of different parts of a system working in harmony, 
and is an ubiquitous phenomenon that has been observed in various branches of sciences 
ranging from physical systems to chemical and biological systems with gain and loss 
\cite{nonlinear1,Pikovsky2001,Kiss1676,Buzsaki1926,Repp2013}. 
Together with other nonlinear effects and beyond, it increases complexity of nature and organizes 
things at higher levels \cite{Anderson}. Synchronization was first discovered by Christian Huygens in 
1665 \cite{Huygens1}. He found that two pendulum clocks hanged side by side would soon swing with the 
same frequency and $180^\circ$ out of phase regardless their initial conditions as long as their intrinsic
frequencies are not too different from each other and their coupling strengths are not too weak. 
This completely out of phase synchronized motion is very robust against the external disturbances. 
Since then, our understanding of synchronization has been greatly advanced.
	 
Two coupled nonlinear oscillators in currently known synchronizations oscillate with the same frequency, 
but can have different constant relative phases \cite{nonlinear1,Varela2001,Huygens1,KuramotoModel}. 
They are relative simple and can be characterized by the frequency and their relative phase. 
For more exotic synchronizations, one needs to couple many nonlinear oscillators as a cluster or a
network \cite{Huang2006} that are commonly described by the Kuramoto model \cite{KuramotoModel}. 
As summarized by Matheny and co-workers \cite{Mathenyeaav7932}, the simplest synchronizations 
of many oscillators are that all oscillators have the same phase, or a few fixed relative phases. 
The relative phases of synchronized oscillator network can even form a complicated static pattern. 
Sometimes, a network can fragment into several clusters, and motions of oscillators in 
each cluster are synchronized with their own static phase pattern. In a word, the patterns 
of phase difference among oscillators in known synchronizations are static and do not 
change with time no matter in coupled two oscillators or in an oscillator network.  

\begin{figure}
\includegraphics[width=0.93\columnwidth]{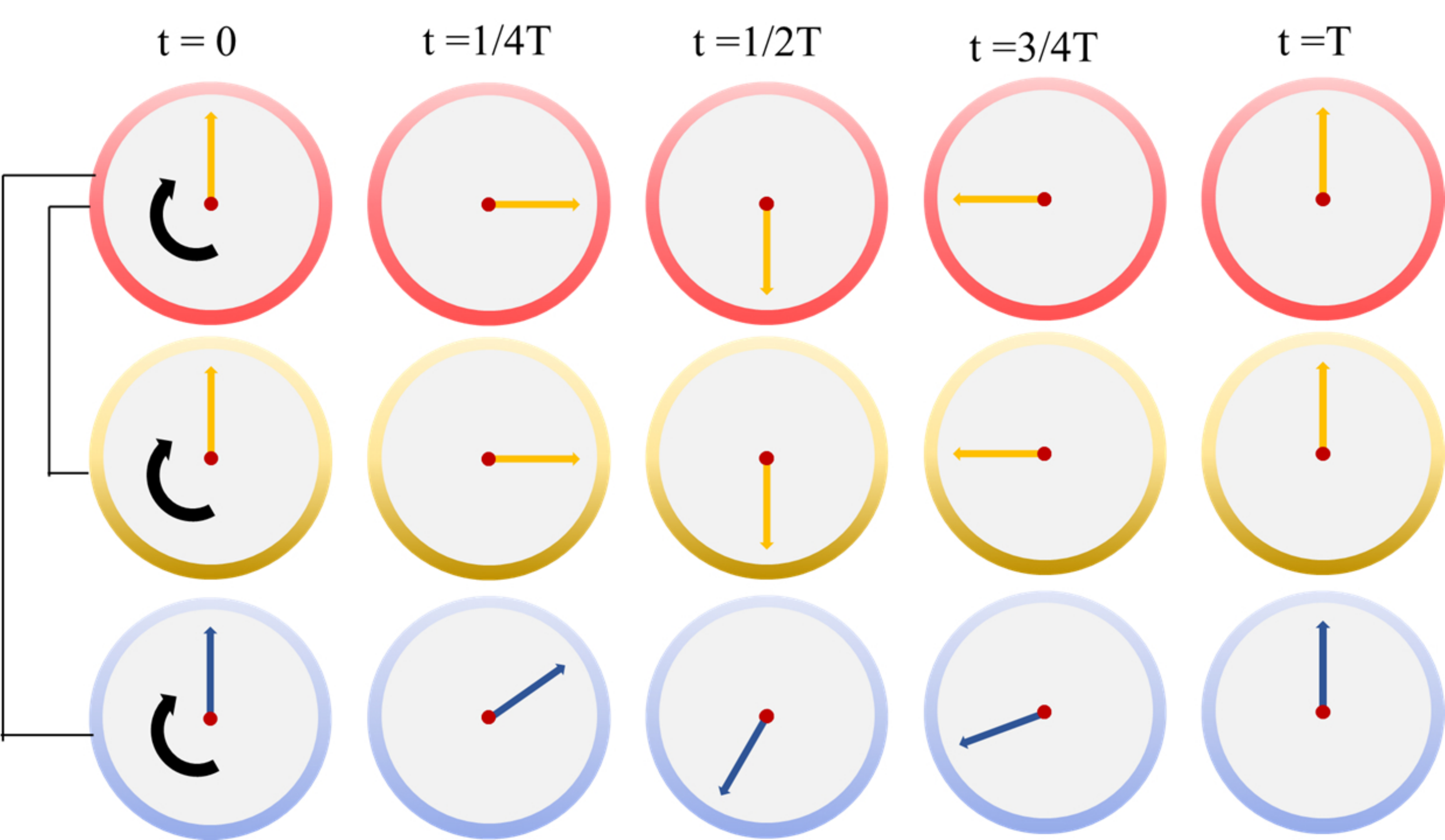}
\caption{Illustration of two types of synchronizations. The red and yellow clocks illustrate 
a conventional in-phase synchronization. Two clocks point to the exactly the same position at 
all times. The red and the blue clocks illustrate a dancing synchronization. 
Two clocks point to 12 o'clock at $t=0$ and complete a cycle exactly in 12 hours. 
In between, the two clocks point to different directions at most of times.}
\label{fig1}
\end{figure}

Spin torque nano-oscillators (STNOs) are important nonlinear oscillators in magnetics. 
STNOs \cite{background1,background2} are self-sustained oscillations driven by  
current generated spin-transfer torque (STT) \cite{SLONCZEWSKI1996L1,Berger1996}. 
Self-sustained oscillations are a well-known nonlinear phenomenon widely existing in 
systems with gain and loss \cite{nonlinear1,qniu,xrw}. STNO is an active research topic in academia 
and industry because of their exotic applications in nano-technology such as microwave generation 
at nano-meter-scale that is crucial for microwave-assisted recording \cite{zzsun1,zzsun2}. 
Output power is an important issue in STNOs \cite{Slavin2009} because microwave power from a single 
STNO is of order of pico-watts due to its tiny size \cite{Kiselev2003}. One promising way of increasing 
the output microwave power is through an in-phase synchronization of many STNOs \cite{Kaka2005,Grollier2006}. 
Several STNOs can be coupled by static magnetic interaction \cite{Chen2016,Belanovsky2012,Huang2013}.  
This coupling is effective only when two STNOs are separated within a few nanometers that limits 
possible number of STNOs in synchronization. Coupling between STNOs through spin waves is order of 
magnitudes larger than that by static magnetic interaction \cite{Dumas2014,Sani2013,Ruotolo2009,
Slavin2006,Slavin2009,Pufall2006,Puliafito2014}. Like other nonlinear systems, various aspects of 
coupled STNOs have been extensively studies, such as the intrinsic mutual phase-locking 
\cite{2005Natur.437..389K,doi:10.1063/1.3278602,refId0,2017NatCo...815825L}, STNOs due to vortex 
state \cite{PhysRevB.80.054412} and the fractional synchronization \cite{PhysRevLett.105.104101}. 
The temperature \cite{PhysRevB.78.092401} and external field \cite{PhysRevB.78.024409} have 
been used to control the frequency, linewidth of STNOs, as well as synchronization.
 
In this study, we report a new type of synchronization of two STNOs coupled by spin waves. 
In the new synchornization the relative phase of two oscillators varies periodically with time, 
instead of being a constant. Such an exotic synchronization is termed {\it dancing synchronisation}. 
Let us use the motion of two coupled clocks, shown in Fig. \ref{fig1}, to explain the 
differences between conventional synchronizations and the dancing synchronizations. 
The red clock (the first row) is in synchronization with both the yellow clock 
(the second row) and the blue clock (the third row) with the same periods, say 12 hours. 
The first and second rows (red and yellow clocks) illustrate several moments of two clocks 
in a conventional synchronization in which two clocks are in phase (always pointing to the same 
direction at all times). The first and third rows (red and blue clocks) schematically 
illustrate relative phases of the two clocks in a dancing synchronization where, within one 
period, the blue clock rotates slower than the red clock in the first and the third phases of 
the period, but faster than the red clock in the second and the last phases of their period. 
The distinct difference of the dancing synchronization from the known ones is that the relative phase 
of the red and blue clocks varies periodically with the synchronized frequency. 

The paper is organized as follows. Section II includes model description of two coupled STNOs, 
methodology, and the demonstration of the dancing synchronization. 
Section III shows that dancing synchronization is universal and exists in well-known complex 
amplitude nonlinear oscillators and the Van der Pol oscillators when there are both reactive 
and dissipative couplings. Then main results are summarised. 

\section{Dancing synchronization in coupled STNOs}
	
\subsection{Model and Methodology}

\begin{figure}
\includegraphics[width=1\columnwidth]{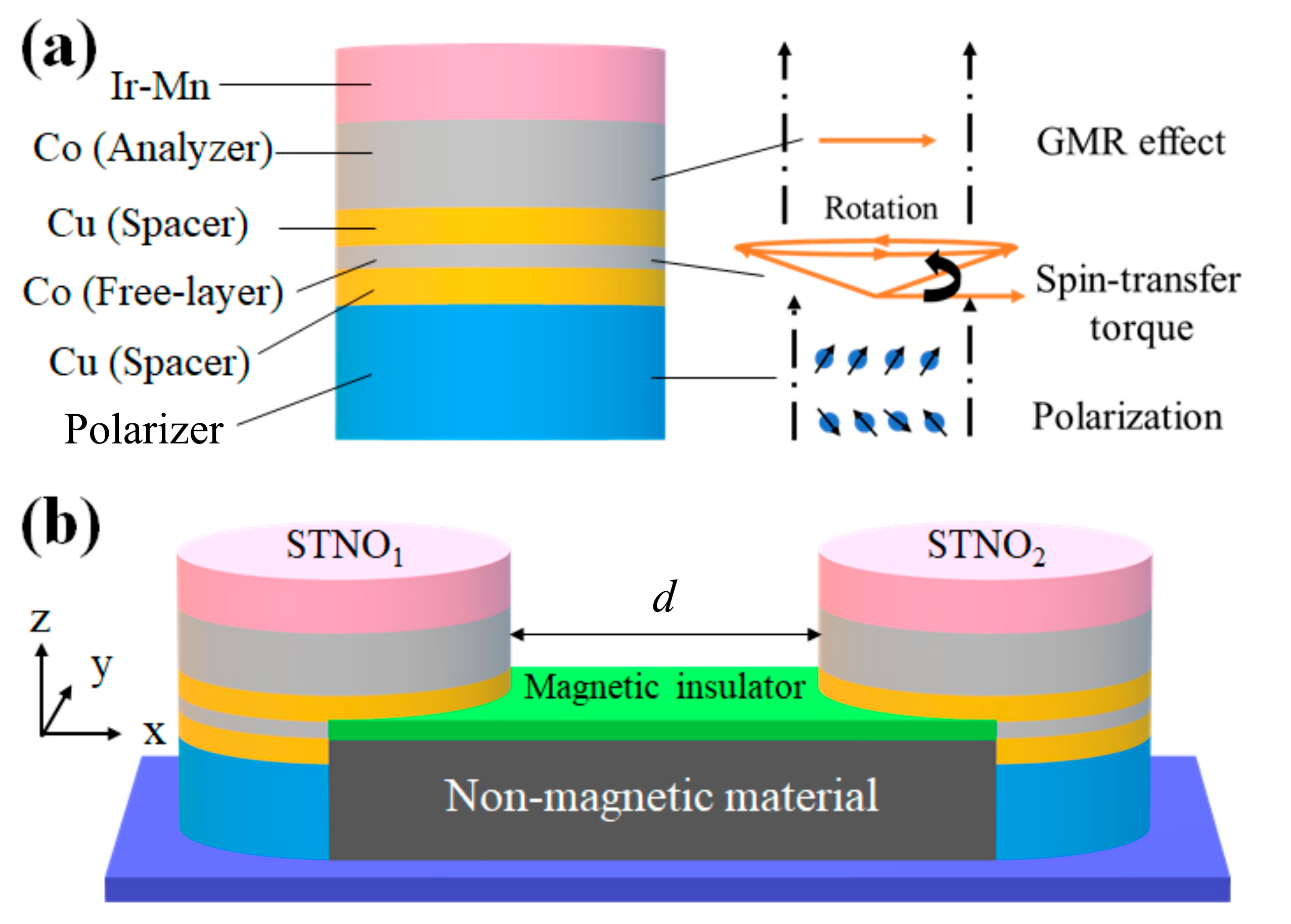}
\caption{Sketch of the model. (a) A typical structure of spin valve in an STNO. 
Self-sustained precession of spins in free-layer is the result of a limit-cycle solution of 
the LLG equation under the spin-transfer torque from spin polarized current that, in turn, 
is obtained by passing current through the polarizer layer. 
(b) Two STNOs connected by a magnetic insulating film are coupled by spin waves in the film. 
$d$ is the distance between two STNOs.} 
\label{fig2}
\end{figure}

Our model, as shown in Fig.~\ref{fig2}, consists of two nano-pillar STNOs coupled through spin waves 
in the magnetic insulating layer physically connected with STNOs. Each STNO is made from magnetic 
multilayer as shown in Fig.~\ref{fig2}(a), which consists of a polarizer of a perpendicularly 
magnetized layer (e.g. $\mathrm{Pt/(Co/Pt)_5}$) to generate spin polarized current; a free layer with 
in-plane magnetization on the top of the polarizer separated by either a nonmagnetic metal such as Cu 
or nonmagnetic insulator such as MgO. Under the STT due to the spin-polarized current from the 
polarizer, the spins in the free layer undergo a self-sustained precession. The self-sustained 
precession can be detected through tunnelling magneto-resistance \cite{Valet1993,Moodera1999} of the 
analyzer on the top of free layer separated by another nonmagnetic layer such as a thin Cu film. 
The analyzer is a thick ferromagnetic film whose magnetization is pinned by an anti-ferromagnetic 
layer (e.g. Ir-Mn) such that self-sustained magnetization precession of free layer can generate an 
oscillatory voltage between the top and bottom layer of the whole nano-pillar shown in the figure. 
Two STNOs have a nominal size of $70\,{\rm nm}\times 60\,{\rm nm}$, and free-layer thickness is of $3 \,\rm nm$. 
The free layer is assumed to be made of Co with saturation magnetization of $M_{\rm s,Co}=886\,\rm kA/m$, 
magnetic anisotropy coefficient of $K=4453\,\rm J/m^3$ (parallel to the line from the center of the 
left STNO to the center of the right STNO), exchange stiffness constant of $A_{\rm Co}=25\,\rm pJ/m$, 
Gilbert damping constant of $\alpha=0.02$ \cite{Houssameddine2007}. Our two STNOs have a slightly 
different spin polarization ($P$) of $P_1=0.38$ for the left STNO and $P_2=0.44$ for the right one. 
The intrinsic oscillation frequencies of the two isolated STNOs under current density of 
$1.435\times10^7\,\rm A/cm^2$ are $9.87\,\rm GHz$ and $10.20\,\rm GHz$, respectively. 
A Yttrium iron garnet (YIG) film of thickness $3\,$nm connects two STNOs as shown in Fig.~\ref{fig2}. 
The material parameters of YIG are $A_{\rm YIG}=4.2\,\rm pJ/m$ and $K_{\rm YIG}=754~ \,\rm J/m^3$  
\cite{Sun2013}. The interface (between the YIG film and STNOs) exchange coupling is 
assumed to be $A_{\rm eff}=2A_{\rm Co}A_{\rm YIG}/(A_{\rm Co}+A_{\rm YIG})$ \cite{ommff}. 
Thus, two STNOs couple through spin waves in the YIG film generated by the STNOs \cite{Dumas2014,
Sani2013,Ruotolo2009,Slavin2006,Slavin2009}, as well as static magnetic interaction 
\cite{Chen2016,Belanovsky2012,Huang2013}. 	

\begin{figure}
\includegraphics[width=\columnwidth]{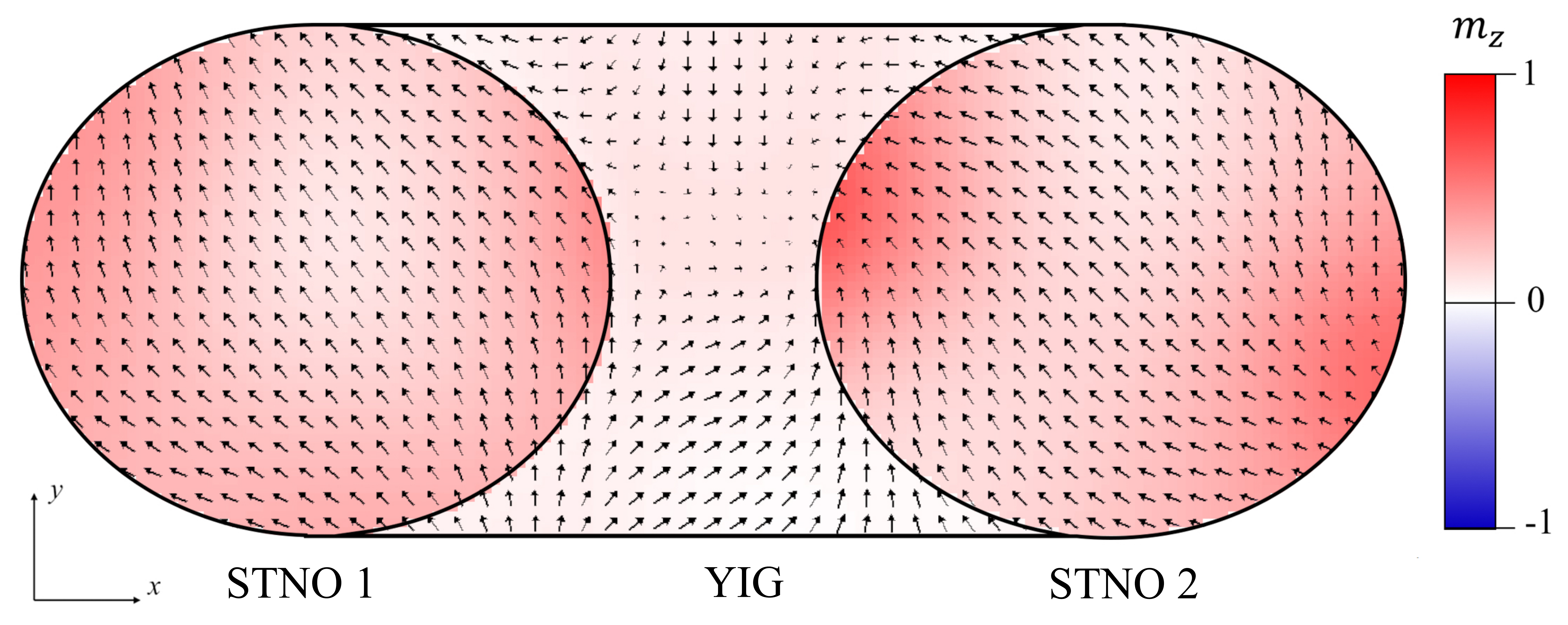}
\caption{A snapshot of spin distribution of system in synchronization. 
The arrows denote the direction of the in-plane component of magnetization and 
the color encodes the the information of $m_z$.}\label{fig3}
\end{figure}

Spin precession in STNO-free-layers will generate and modify spin waves in the YIG film 
such that two STNOs can interact with each other through the exchange of spin waves. 
This spin wave mediated coupling is much stronger \cite{Ruotolo2009} than the direct 
magnetic-dipole interactions between two STNOs when they are close to each other. 
The STNO separation, material parameters, and the applied electrical current can be used to control 
the effective coupling of STNOs. We investigate the spin dynamics of the hybrid structure 
consisting of free layers of STNOs and the YIG film under the injection of spin polarized currents. 
The current density has a non-zero value only within the free-layers of STNOs. 
The thermal effect, the field generated from the analyzer layer (not show) and the field induced by 
charge current are ignored. Spin dynamics of the system is governed by the Landau-Lifshitz-Gilbert 
(LLG) equation,
\begin{equation}
\begin{split}
\frac{\mathrm{d}\mathbf{m}}{\mathrm{d}t}=&-\gamma\mathbf{m}\times\mathbf{H}_{\rm{eff}}+\alpha(\mathbf{m}\times
\frac{\mathrm{d}\mathbf{m}}{\mathrm{d}t})\\   &+{a}(\mathbf{m} \times\mathbf{m}_{\rm p}\times\mathbf{m}),
\end{split}
\label{llg}
\end{equation}
where $\mathbf{m}$, $\gamma$, $t$, and $\mathbf{H}_{\rm{eff}}$ are respectively the unit vector of the 
magnetization, gyromagnetic ratio, the time, and the effective magnetic field, $\mathbf{H}_{\rm{eff}}=
\frac{2A}{\mu_0M_{\rm s}} \nabla^2\mathbf{m}+\frac{2K}{\mu_0M_{\rm s}}m_z\hat z+\mathbf{H}_{\rm d}$ that includes the 
exchange field, the anisotropic field, and the demagnetizing field $\mathbf{H}_{\rm d}$.
Coefficient $a=|\frac{\hbar}{\mu_0e}|\frac{J}{dM_{\rm s}}\frac{P\lambda^2}{(\lambda^2+1)+
(\lambda^2-1)(\mathbf{m}\cdot\mathbf{m}_{\rm p})}$ describes the Slonczewski torque, where $\hbar$, $d$, 
$M_{\rm s}$, $J$, $e$, $\mu_0$, and $P$ are the reduced Planck constant, the thickness of free 
layer, the saturation magnetization of the free layer, the charge current density, the electron 
charge, the vacuum permeability, and the polarization of the charge current, respectively. 
Under a proper spin polarized current, the spins in the free layer undergo a self-sustained precession. 
Eq.~\eqref{llg} for the whole hybrid system of YIG film and free layers in STNOs is numerically solved 
by using the OOMMF (Object Oriented MicroMagnetic Framework) \cite{ommff}. To balance the speed and 
accuracy, the cell size used in this study is of $1\,$nm$\times1\,$nm$\times3\,$nm.

Initially, spins of the left STNO are all along the $x$-direction, and all spins of the right STNO 
are in the $yz$-plane and $45^{\circ}$ away from the $z$-axis. Under an electric current density of 
$1.435\times10^7\,\rm A/cm^2$, two STONs are synchronized after few nanoseconds when the distance 
between STNOs is $d=22\,$nm. Figure~\ref{fig3} is a typical snapshot of spin configuration of two 
STNOs in the synchronization where spins in both STNOs and YIG do not align along the same direction 
even in the synchronised state because of the edge and interface effect. 

\subsection{Coupling length of the STNOs with spin-wave in YIG}

\begin{figure}
\includegraphics[width=\columnwidth]{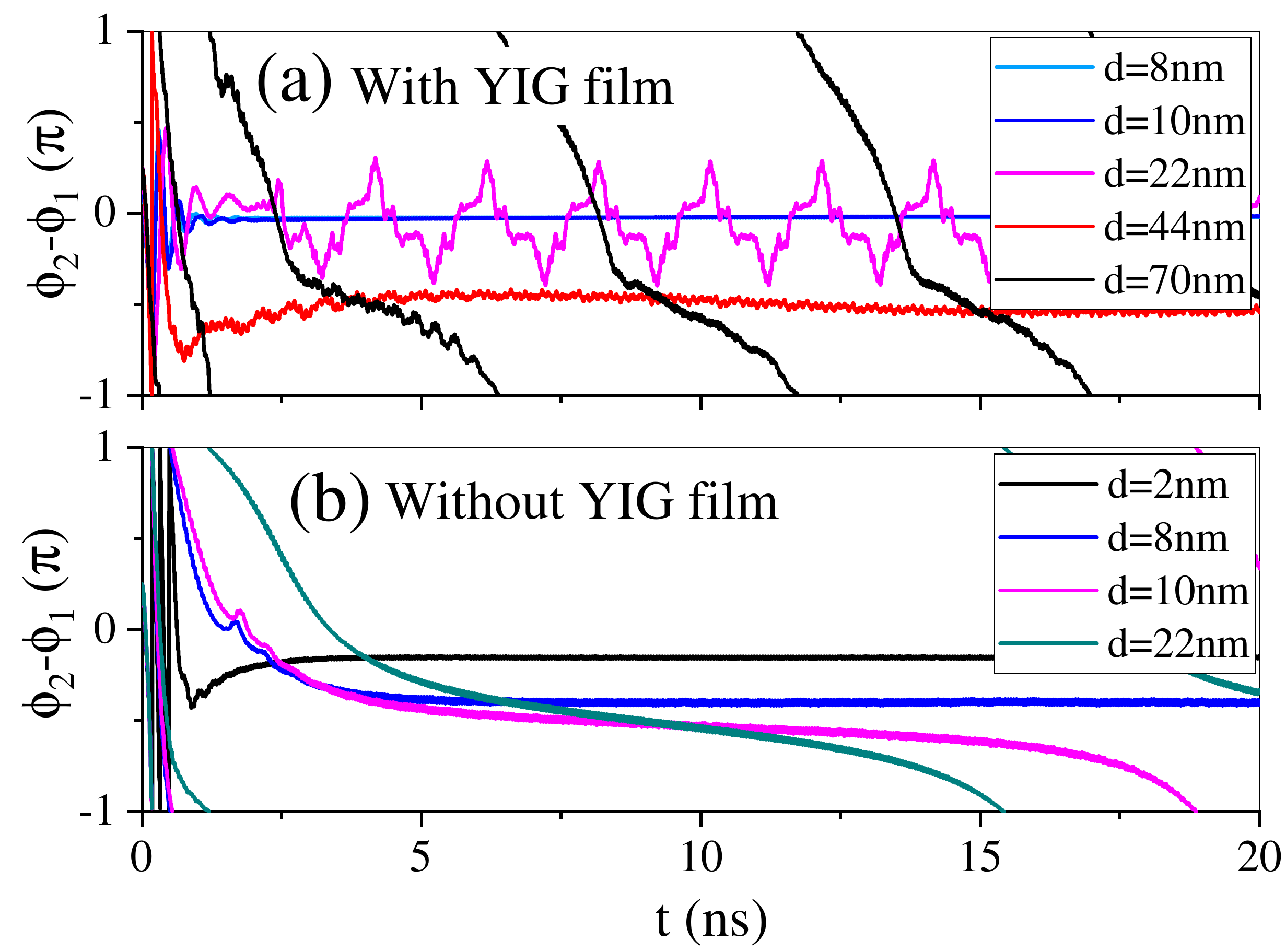}
\caption{Time evolution of phase differences with spin-wave coupling (a) and with only dipolar 
coupling (b) for various distances and a fixed charge current density of $1.435\times10^7\,\rm A/cm^2$.}
\label{fig4}
\end{figure}

\begin{figure*}[htbp]
\includegraphics[width=2\columnwidth]{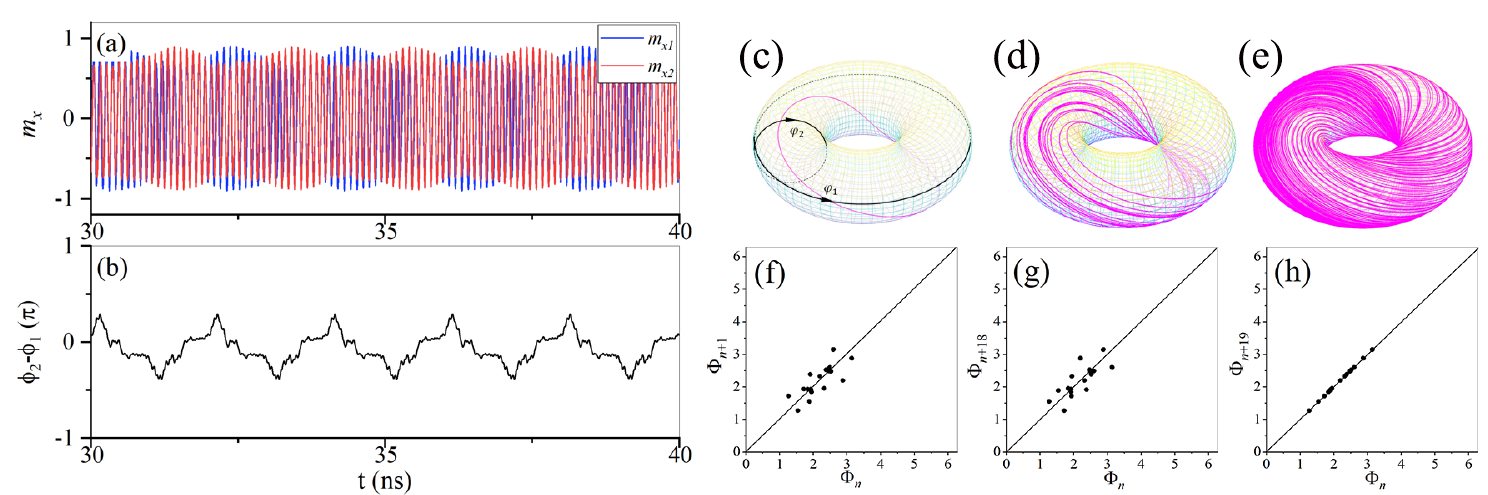}
\caption{(a) Time evolutions of $m_{1x}(t)$ (the blue curve) and $m_{2x}(t)$ (the red curve) in a 
dancing synchronization: $m_{1x}(t)$ and $m_{2x}(t)$ show a fast and a slow motion (at nanoseconds). 
The relative phase of the red and the blue curves varies with a much longer common period. 
(b) Time evolution of phase difference in the dancing synchronization. 
The common long period is about $2\,$ns, much longer than a tenth nanosecond oscillation.
(c)-(e) Phase trajectories $\phi_2(\phi_1)$ of two STNOs on $\phi_1\phi_2$-torus ($\phi_1$ for the 
large circle and $\phi_2$ for the smaller one), under a current density of $1.435\times10^7\,\rm A/cm^2$; 
(c) is for a conventional in phase synchronization when the distance is $10\,$nm ; (d) is for the 
dancing synchronization in which $\phi_2(\phi_1)$ return to its starting point after 19 turns 
when the distance is $22\,$nm and (e) is for a non-synchronized state in which $\phi_2(\phi_1)$ never 
closes when the distance is $70\,$nm. (f)-(h) $\Phi_n=\phi_2(\phi_1=\pi)$ is the value of $\phi_2$ in 
the Poincare maps. (f-h) are respectively for $\Phi_n$ vs. $\Phi_{n+1}$, $\Phi_n$ vs. 
$\Phi_{n+18}$, and $\Phi_n$ vs. $\Phi_{n+19}$.} \label{fig5}
\end{figure*}

We first study the coupling distance of the two STNOs through the spin waves in the YIG film. 
We use OOMMF to simulate two identical systems described above except that one of them 
does not have the YIG film such that two STNOs couple with each other by dipolar field. 
Thus, one can attribute difference of two system to the spin wave mediated coupling. 
To see different behaviour of the two systems, we collect time evolution data of the  
average magnetization $\mathbf{m}_i(t)$ of two STNOs,  where $i=1,2$ label the two STNOs.
The angles of in-plane component of $\mathbf{m}_i(t)$ with the $x$-axis are denoted as $\phi_i(t)$. 
The time dependence of phase difference $\phi_{2}(t)-\phi_{1}(t)$ can tell synchronizations 
from non-synchronisations. $\phi_{2}(t)-\phi_{1}(t)$ varies over 2$\pi$ range in a 
non-synchronized motion while it is a constant in a conventional synchronization. 
Our OOMMF simulation results are shown in Fig.~\ref{fig4}(a) for system with YIG film, 
and in Fig.~\ref{fig4}(b) for system without YIG film. Indeed, both non-synchronisations 
[for $d=70\,$nm in Fig.~\ref{fig4}(a) and $d=10,22\,$nm in Fig.~\ref{fig4}(b)] and conventional synchronizations 
[for $d=8,10,44\,$nm in Fig.~\ref{fig4}(a) and $d=2,8\,$nm in Fig.~\ref{fig4}(b)] can be clearly identified. 
Interestingly, a periodically oscillating $\phi_{2}(t)-\phi_{1}(t)$ with an amplitude of $60^\circ$ 
appears at $d=22\,$nm in the case that two STNOs are coupled by both dipolar field as well as 
by the spin waves due to YIG film. This is exactly the dancing synchronization discussed early. 
Without the spin waves, such a synchornization was not observed [Fig.~\ref{fig4}(b)].   
Therefore, results in Fig.~\ref{fig4} demonstrate not only that coupling distance 
between two STNOs by spin waves becomes much longer ($44\,$nm) than that ($8\,$nm) by 
dipolar field, but also it can induce a new type of synchronization never observed 
before. 
Below, we will examine this new synchronization more closely. 

\subsection{Dancing synchronization}
	
For the dancing synchronization at $d=22\,$nm and under current density of 
$1.435\times10^7\,\rm A/cm^2$, we plot the time evolutions of $m_{1x}(t)$ (the blue 
curve) and $m_{2x}(t)$ (the red curve), $x$-components of average magnetization of 
free layer in the left and the right STNOs,  respectively, in Fig.~\ref{fig5}(a). 
Two curves are periodic with the same period, but have different shapes, i.e. $m_{\alpha 
x}(t)=m_{\alpha x}(t+nT)$ ($\alpha=1,2$), where $T$ is the period and $n$ is an 
arbitrary integer. For example, within one common period, both $m_{1x}(t)$ and $m_{2x}(t)$ 
oscillate 19 times with different amplitudes before returning to their initial values. 
This phenomenon is different from the conventional in-phase synchronization, 
where time evolutions of $m_{1x}(t)$ and $m_{2x}(t)$ either overlap completely 
with each other or differ by a fixed lag. Fig.~\ref{fig5}(b) plots time 
evolution of the phase difference $\phi_{2}(t)-\phi_{1}(t)$ of the two STNOs. 
Clearly, $\phi_{2}(t)-\phi_{1}(t)$ oscillates periodically with an amplitude 
of about $\pi/3$ and a period of $2~ns$. This is different from all known 
synchronizations where $\phi_{2}(t)-\phi_{1}(t)$ is a constant. 
Because of this periodical variation of relative phase of the two STNOs that 
is reminiscent of two partners dancing in rhymes with different arm movements, 
we term this observed new synchronization of {\it dancing synchronization}. 

\begin{figure*}[htbp]
\includegraphics[width=2\columnwidth]{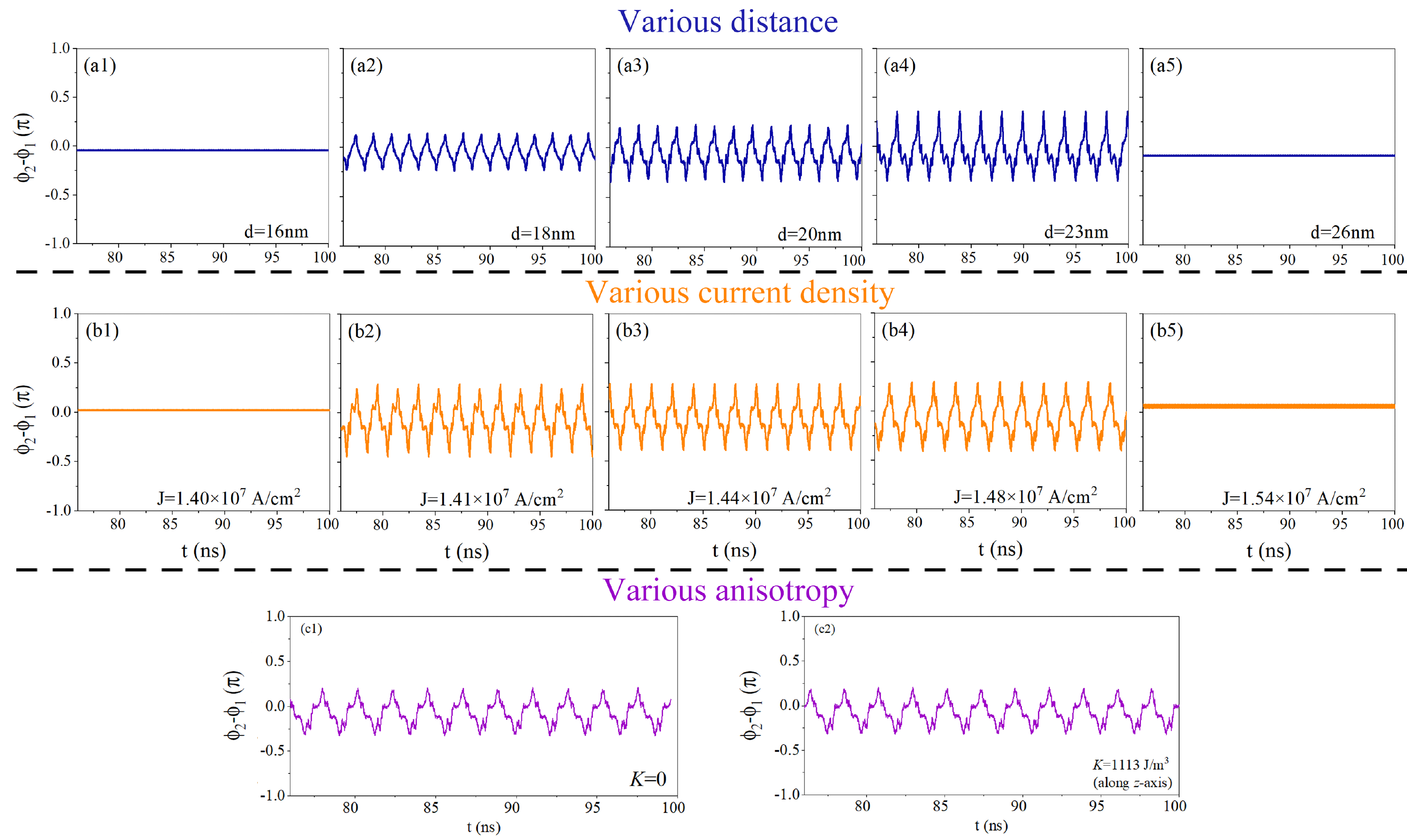}
\caption{Time evolution of phase differences for various distance $d$ at a 
fixed charge current $J=1.435\times10^{7}\,\rm A/cm^2$; for various charge current 
$J$ at a fixed distance $d=22\,$nm and for different conduction of the magnetic 
anisotropy at fixed $d=22\,$nm and $J=1.435\times10^{7}\,\rm A/cm^2$. 
All other unmentioned parameters are the same as those used in Fig.~\ref{fig5}.} \label{fig6}
\end{figure*}

To further prove the dancing synchronization of Fig.~\ref{fig5}(a), we plot  
trajectory $\phi_2(\phi_1)$ on the $\phi_1\phi_2$-torus as shown in Fig.~\ref{fig5}(c-e). 
In a conventional synchronization where $\phi_2(t)-\phi_1(t)=const.$, $\phi_2(t)$ and $\phi_1(t)$ change by 
$2\pi$ simultaneously so that $\phi_2(\phi_1)$ is a simple one-turn closed curve as shown in Fig.~\ref{fig5}(c).
This is the case when the distance between the two STNOs is $10\,$nm under a current density of $1.435\times10^{7}\,\rm A/cm^2$.
The case of $d=22\,$nm at the same current density is fundamentally different as shown in Fig.~\ref{fig5}(b). 
$\phi_2(t)-\phi_1(t)$ is not a constant, but varies periodically with a longer period. The trajectory is still 
a closed curve as shown in Fig.~\ref{fig5}(d) that displays data of Fig.~\ref{fig5}(a) as $\phi_2(\phi_1)$ on 
the $\phi_1\phi_2$-torus. $\phi_2(\phi_1)$ returns to its starting point after 19 turns.
If $\phi_1(t)$ and $\phi_2(t)$ either are neither periodic nor have a common period, the trajectory will not be 
a closed curve and will fill up the $\phi_1\phi_2$-torus, as shown in Fig.~\ref{fig5}(e) that is 
the motion of the two STNOs for $d=70\,$nm and under a current density of $1.435\times10^7\,\rm A/cm^2$. 

One can further confirm the dancing synchronization of two STNOs in Fig.~\ref{fig5}(a) via the Poincare maps. 
In the map, $\Phi_n$ is defined as angle $\phi_2$ modulo $2\pi$ when $\phi_1 =(2n-1)\pi$, i.e. 
\{$\Phi_n=\phi_2(\phi_1=(2n-1)\pi) \ modulo \ 2\pi| n=1,2,\ldots$\}. 
$\Phi_n$ can be grouped into various sets such as $\{(\Phi_n,\Phi_{n+1})|n=1,2,\ldots\}$, or 
$\{(\Phi_n,\Phi_{n+18})|n=1,2,\ldots\}$, or $\{(\Phi_n,\Phi_{n+19})|n=1,2,\ldots\}$. 
These three sets are plotted in Fig.~\ref{fig5}(f-h) where the $x$-axis is for $\Phi_{n}$ and the $y$-axis 
for $\Phi_{n+N}$, $n=1$, 2, 3, ... $\{(\Phi_n,\Phi_{n+N})|n=1,2,\ldots\}$ 
fall onto the line of $\Phi_{n+N}=\Phi_{n}$ if $\phi_1$ and $\phi_2$ have the common period of $N$ turns. 
This is exactly the case here with $N=19$ as shown in Fig.~\ref{fig5}(h). As a comparison, sets with 
$N=1$, and 18 are off the straight line as shown in Fig.~\ref{fig5}(f) and Fig.~\ref{fig5}(g).  

\subsection{Robustness of the dancing synchronization}
The observed dancing synchronization is very robust, and can exist in a finite region in the parameter space.  
For example, Fig.~\ref{fig6}(a1$\sim$a5) shows the time evolution of $\phi_{2}(t)-\phi_{1}(t)$ for various 
$d$ at a fixed current density of $J=1.435\times10^7\,\rm A/cm^2$ while all other parameters keep the same as those 
for Fig.~\ref{fig5}. Clearly, the dancing synchronization, featured by the periodic variation of $\phi_{2}(t)
-\phi_{1}(t)$, occurs in the window of $d=18\sim23\,$nm. Similarly, we observe the dancing synchronization at 
fixed $d=22\,$nm in the current density window of $J=1.41\sim1.48\times10^{7}\,\rm A/cm^2$ while all other parameters 
keep the same as those for Fig.~\ref{fig5}, as shown in Fig.~\ref{fig6}(b1$\sim$b5) where $\phi_{2}(t)
-\phi_{1}(t)$ in Fig.~\ref{fig6}(b2-b4) vary periodically. Moreover, as shown in Fig.~\ref{fig6}(c2) at a fixed $d=22\,$nm, 
$J=1.435\times10^7\,\rm A/cm^2$, the dancing synchronization occurs when the magnetic anisotropy direction as 
well as its magnitude vary. Interestingly, the dancing synchronization exists even in the absence of the 
anisotropy as shown in Fig.~\ref{fig6}(c1).

\begin{figure}
\includegraphics[width=\columnwidth]{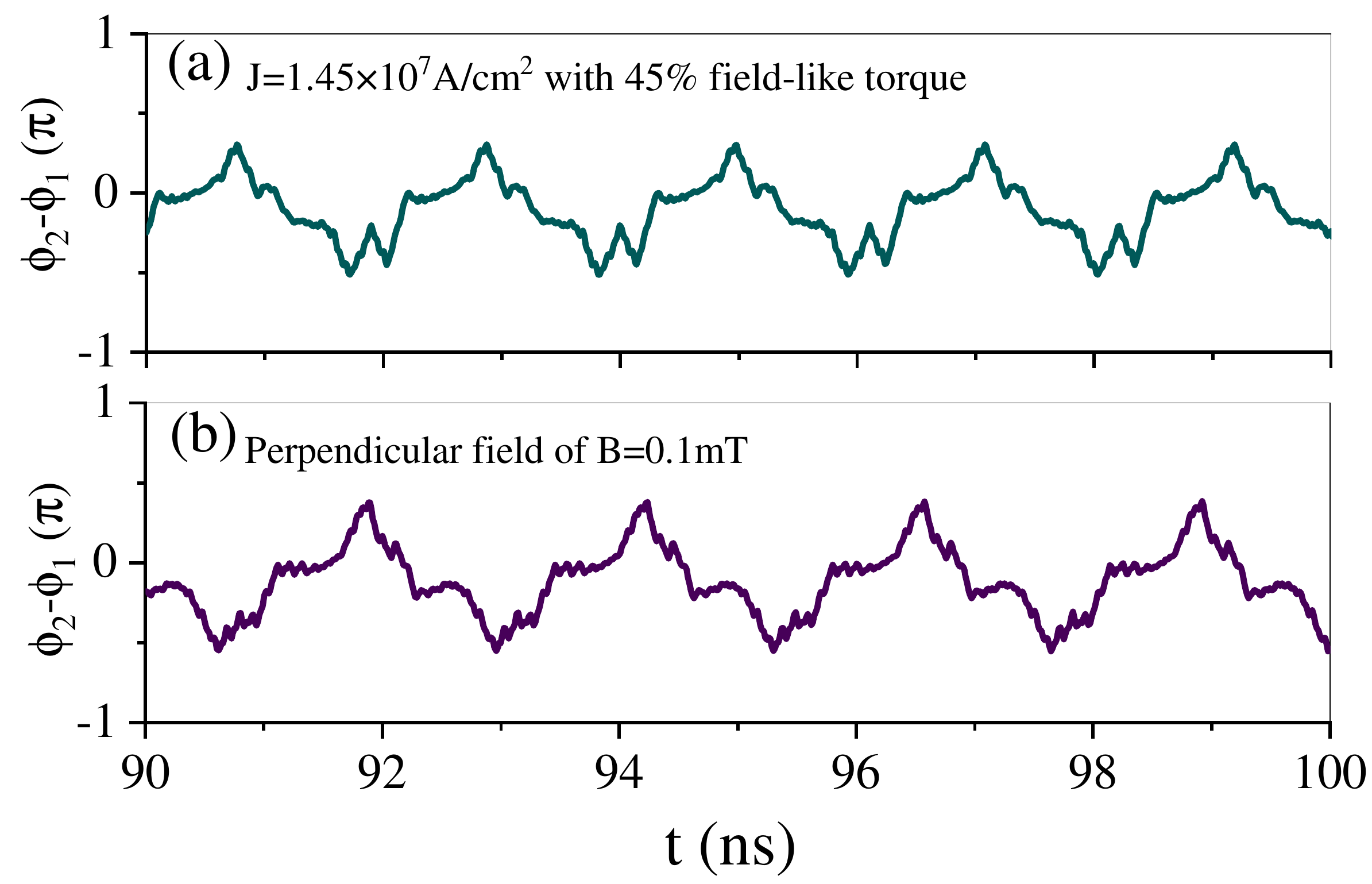}
\caption{Time evolution of phase differences at a fixed distance $d=22\,$nm and under 
a current density of $1.45\times10^7\,\rm A/cm^2$ with 45\% field-like torque (a); or under  
an external perpendicular magnetic field of 0.1 mT (b). 
All other unmentioned parameters are the same as those used in Fig.~\ref{fig5}.}\label{fig7}
\end{figure}

A natural question is whether the dancing synchronization can still survive when the so-called field-like 
torque is included in Eq.~\eqref{llg}. The answer is yes as shown in Fig.~\ref{fig7}(a) for $d=22\,$nm and 
under a current density of $1.45\times10^7\,\rm A/cm^2$ with 45\% field-like torque. 
The torque modifies slightly the details of the synchronization.
The dancing synchronization is still observed even when an additional external magnetic field up to $0.3\,$mT 
along the $z$-axis is applied, as shown in Fig.~\ref{fig7}(b) for $0.1\,$mT. 
These results demonstrate the robustness of the dancing synchronization against parameters and 
different types of torques.

\begin{figure}
\includegraphics[width=\columnwidth]{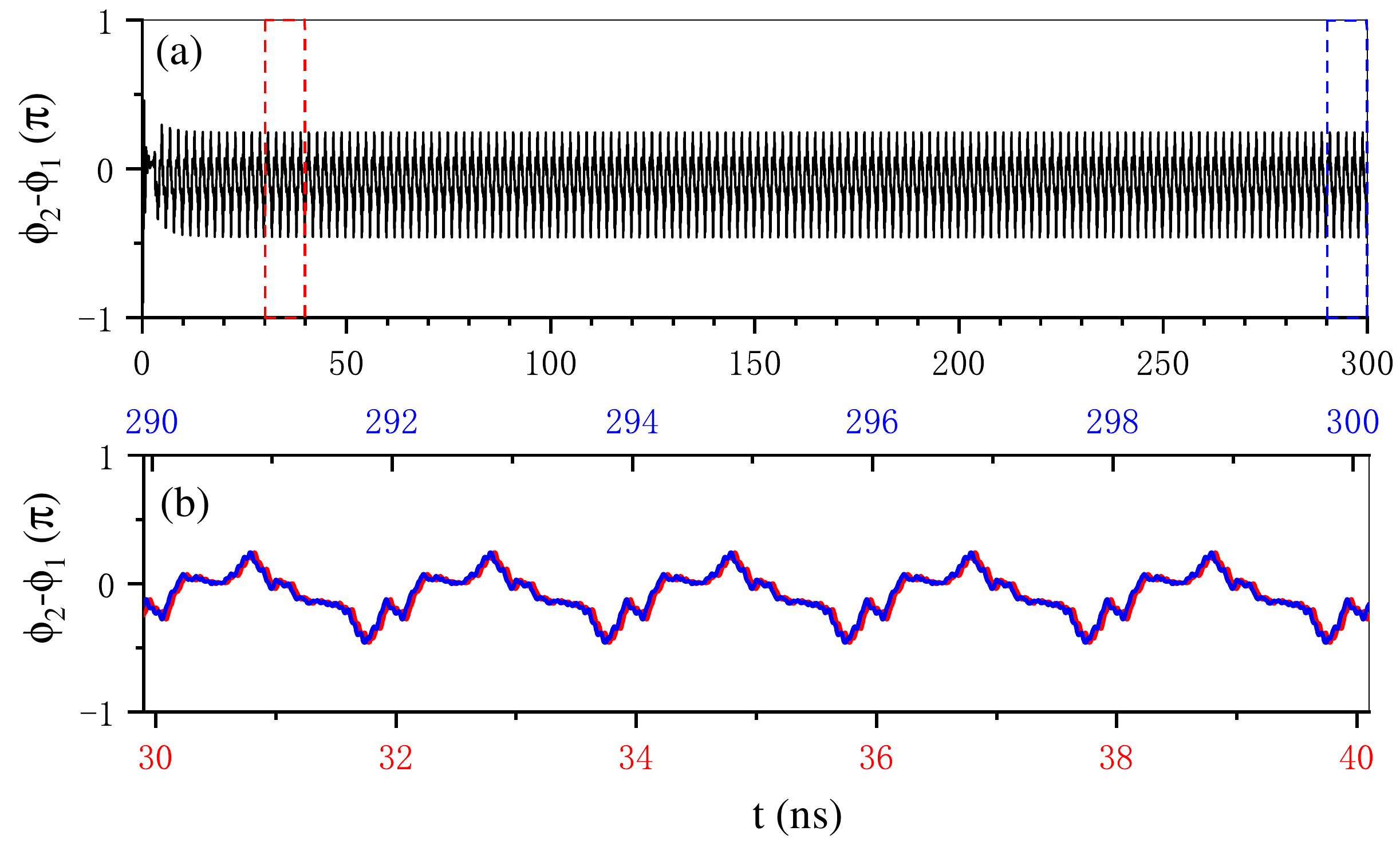}
\caption{(a) A $300\,$ns long evolution of phase differences
at $d=22\,$nm and $K_{YIG}=0$ under current density of $1.435\times10^7\,\rm A/cm^2$. 
(b) Zooming in of the evolutions in $t=30\sim40\,$ns (bottom axis and red curve) and in $t=290\sim300\,$ns (top axis and blue curve).
Two curves overlap with each other, showing no sign of a transient motion.}
\label{fig8}
\end{figure}

The observed dancing synchronization is not a transient process. This can be verified by a 
much longer micromagnetic simulation of 300$\,$ns. 
In this simulation, we set $d=22\,$nm and $K_{YIG}=0$ in order to show that the
dancing synchronization is robust against variation of spin waves 
that glue two STNOs together.
The rest of model parameters are the same as those in Fig. 5.
As shown in Fig.~\ref{fig8}, there 
is no sign that the dancing synchronization changes to another type of motion. 
Evolution of phase difference between $t=290\,$ns and $t=300\,$ns is the same as that between $t=30\,$ns and $t=40\,$ns, and is very similar to Fig.\ref{fig5}(b) with $K_{YIG}\neq 0$. 

\section{Dancing synchronization in toy models}\label{analytical}
Importantly, genuine physics phenomena should be universal. In order to demonstrate that 
the dancing synchronization can also appear in well-known and well-studied popular models, we 
consider two coupled complex variable oscillators \cite{Mathenyeaav7932} and two coupled Van 
der Pol oscillators. The two models have been intensively studied by numerous people before, 
and, to the best of our knowledge, no dancing synchronization has been reported to date. 
A close examine of earlier studies shows that most people use the simple linear reactive coupling (function 
of oscillator position only) between two nonlinear oscillators. Indeed, we did not observe the dancing 
synchronization with only linear reactive or linear dissipative coupling (involving oscillator 
velocities) like previous studies. However, when two nonlinear oscillators couple with each other 
both reactively and dissipatively, dancing synchronization appears. Below, we report our findings. 

\subsection{Dancing synchronization in coupled complex variable oscillators}	
\label{cc}

We first search dancing synchronization in the complex variable oscillation model used by Matheny and 
co-workers \cite{Mathenyeaav7932} who reported various fragmentation synchronizations.
The nonlinear dynamical equations for $n$-complex-variables $A_j(t)$ ($j=1, \ldots n$) read 
\begin{equation}
\begin{split}
\dot{A_j}=&{\lambda}A_j(1-|A_j|)+i(\omega_jA_j+\alpha|A_j|^2A_j) \\
&+i\beta\sum_{k\neq j}^n (A_k-A_j)+{\gamma}\sum_{k\neq j}^n A_k(1-|A_j|),
\label{complex-osc}
\end{split}
\end{equation}
where $\alpha$ is the nodal nonlinearity that couples frequency to amplitude, $\beta$ measures 
the strength of reactive coupling among a pair of oscillators, $\gamma$ is a non-linear coupling. 
Each complex variable $A_j(t)$ stands for an oscillator. The real part of $A_j(t)$ represents a real 
variable which can be observed in the oscillation. Equation \eqref{complex-osc} is often used to 
introduce the concept of synchronization \cite{nonlinear1}. For STNOs, $A_j(t)$ can be $\mathbf{m}_j$. 
The phase of each oscillator $\phi_{j}(t)$ is defined as the argument of $A_j$.
For $\gamma=0$, the model has been used to describe various nonlinear systems including NEMS \cite{Mathenyeaav7932}. 
This model is sometimes called ``a universal model for self-sustained oscillations" in comparison 
to the Kuramoto model \cite{KuramotoModel} widely used to describe ``phase synchronization" 
of coupled oscillators or networks. In Kuramoto model, an oscillator is represented by only 
one real variable.  

\begin{figure*}
\includegraphics[width=1.6\columnwidth]{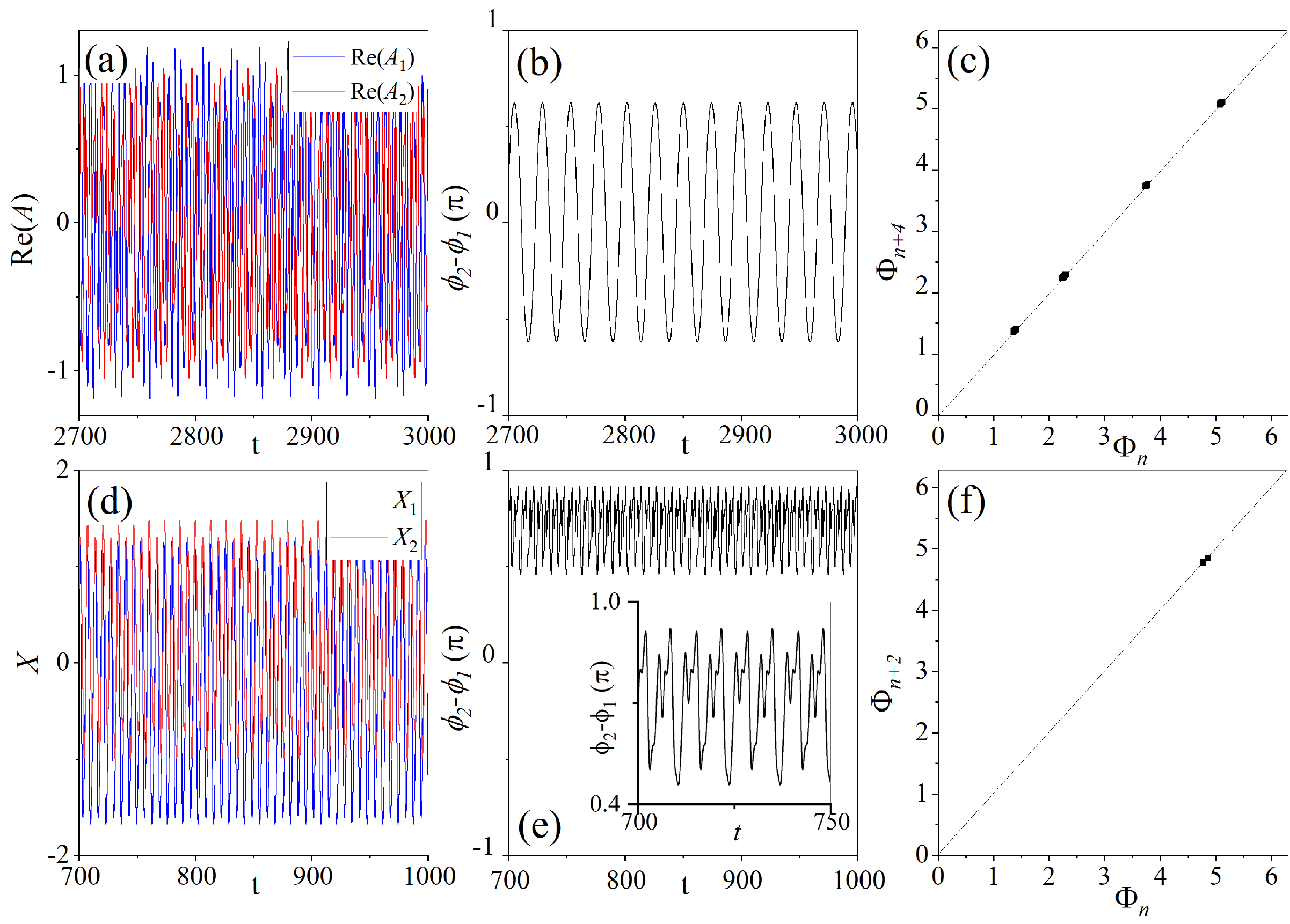}
\caption{Dancing synchronizations in complex amplitude model and Van der Pol model. 
(a), (b) and (c) are time evolution of complex amplitudes, phase difference and 
the Poincare map in complex amplitude model, respectively. 
(d), (e) and (f) are real time trace of two oscillators' amplitudes, the time evolution of phase difference 
, and the Poincare map in the VdP model, respectively. }
\label{fig9}	
\end{figure*}

Various nonlinear phenomena such as self-sustained oscillation and fragmentation synchronizations have been 
obtained from Eq.~\eqref{complex-osc} with $\gamma=0$ \cite{Mathenyeaav7932}, but not the dancing synchronization. 
We show now that the dancing synchronization of two complex-variable oscillators can exist for certain $\gamma\neq 0$. 
The numerical solutions of Eq.~\eqref{complex-osc} from fourth-order Runge Kutta method are plotted in 
Fig.~\ref{fig9}(a) and Fig.~\ref{fig9}(b) for $\gamma=0.01$, $\omega_1=0.5\,$Hz, $\omega_2=0.7\,$Hz, $\alpha=0.59126$, 
$\beta=0.056$, $\lambda=0.01$ with the initial conditions $A_1 (0)=2.51e^{0.16i}$ and $A_2(0)=1.62e^{0.79i}$. 
Similar to the STNOs system, the amplitudes $Re(A_1)$ and $Re(A_2)$, as shown in Fig.~\ref{fig9}(a), oscillate with a 
long common period of $24.75 s$. The phase difference $\phi_{2}(t)-\phi_{1}(t)$ as shown in Fig.~\ref{fig9}(b) is not 
a constant, but varies with the same synchronized period of $24.75\,$s with an amplitude of $0.6\pi$. 
Due to the fact that all nonlinear dynamical systems with gain and loss, the properties of attractors 
do not depend on the initial states. The dancing synchronization is also checked using phase trajectory 
and the Poincare map \{$\Phi_n=\phi_2(\phi_1=(2n-1)\pi) \ modulo \  2\pi| n=1,2,\ldots$\}, and phase trajectory 
$\phi_{2}(\phi_{1})$ is closed after 4 turns on $\phi_1\phi_2$-torus as demonstrated by the points of 
$\{(\Phi_n,\Phi_{n+4})|n=1,2,\ldots\}$ on line $y=x$ in Fig.~\ref{fig9}(c). 

\subsection{Dancing synchronization in two coupled Van der Pol oscillators}
\label{vdp}
	
We have also demonstrated existence of the dancing synchronization in coupled two Van der Pol (VdP) nonlinear oscillators. 
The VdP equation is not only a popular model for demonstrating the self-sustained oscillation in nonlinear 
systems \cite{VdP1,zzsun}, but also realizable by RCL-circuits with a negative differential resistor. 
The standard VdP equation is,
\begin{equation}
\ddot{x_i}+\mu(x_i^2-A_i)\dot{x_i}+\omega_i^2x_i=-f_{i,j\neq i},
\label{m-vdp}
\end{equation}
where $i,j=1,2$ label two oscillators; $\mu>0$ is a parameter measuring energy gain ($x_i^2<A_i$) and energy 
loss ($x_i^2>A_i$). $A_i>0$ specifies the size of energy gain region and is roughly oscillation amplitude. 
$\omega_i$ and $f_{ij}$ describe respectively the oscillatory frequency and the coupling between oscillators $i$ and $j$.
Coupled VdP oscillators have been intensively studied before with either reactive or dissipative 
coupling \cite{Bifurcation-vdp,Delay-vdp}. Interestingly, only conventional synchronizations were 
reported in all earlier studies of coupled VdP oscillators. Here we show that the dancing synchronization 
can appear in coupled VdP oscillators with both reactive and dissipative couplings, 
\begin{equation}
\begin{split}
f_{ij}=\alpha(x_j-x_i)+(j-i)\beta\sqrt{|x_ix_j+\dot{x_i}\dot{x_j}-1|},
\label{coupleterm}
\end{split}
\end{equation}
where the first term is a reactive coupling and the second one is dissipative.
Figure \ref{fig9}(d) are numerical solutions of Eq.~\eqref{m-vdp} from fourth-order Runge 
Kuta method for $\mu=1$, $A_1=A_2=0.5$, $\omega_1=1\,$Hz, $\omega_2=0.98\,$Hz, $\alpha=0.12$, $\beta=0.30$.
The final self-sustained oscillations shown in those figures do not depend on the initial conditions.
Two oscillators have distinguished appearances, but share a common long period of $13.19\,$s.
To see clearly that this is a dancing synchronization, we define 
\begin{equation}
\phi_{j}(t)=\int_{0}^{t}\frac{\dot{x_j}(\tau)\dddot{x_j}(\tau)-\ddot{x_j}(\tau)^2}{\dot{x_j}(\tau)^2-\ddot{x_j}(\tau)^2}\,\mathrm{d}\tau,
\end{equation}	
which is the total winding angle of $(x(t),\dot{x})$ in $x\dot{x}$ phase-plane.
$\phi_{2}-\phi_{1}$ varies periodically with an amplitude of around $0.2\pi$ within the common long period of $13.19\,$s, 
as plotted in Fig.~\ref{fig9}(e). 
Again, the dancing synchronization is checked using phase trajectory and the Poincare map 
\{$\Phi_n=\phi_2(\phi_1=(2n-1)\pi) \ modulo \ 2\pi| n=1,2,\ldots$\}, and phase trajectory 
$\phi_{2}(\phi_{1})$ is closed after 2 turns on $\phi_1\phi_2$-torus as demonstrated by the points of 
$\{(\Phi_n,\Phi_{n+2})|n=1,2,\ldots\}$ on line $y=x$ in Fig.~\ref{fig9}(f).

\subsection{Discussion}
\begin{figure}
\includegraphics[width=\columnwidth]{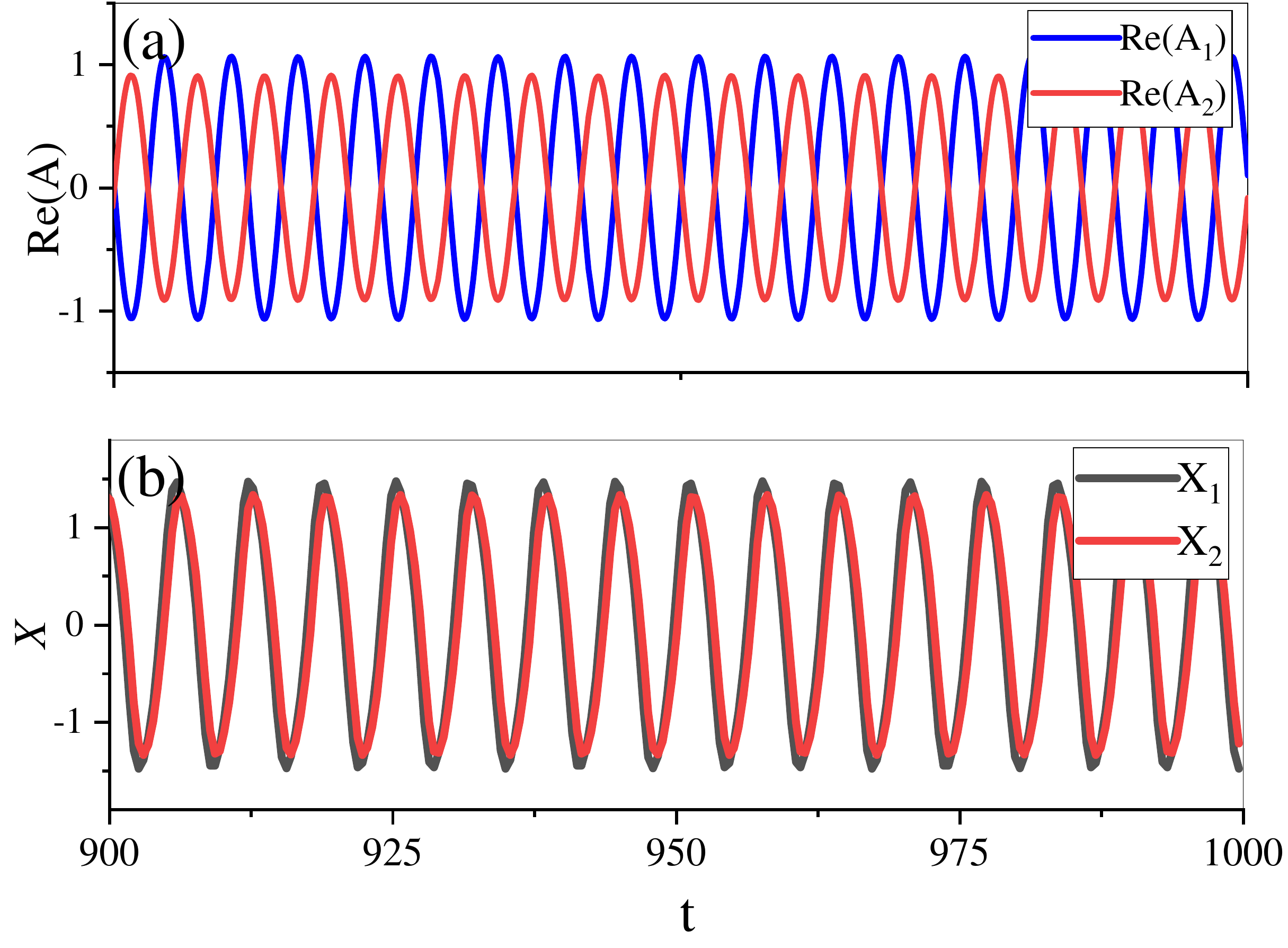}
\caption{Time evolution of two nonlinear oscillators for the complex amplitude model with $\gamma=0$ (a) 
and the Van der Pol model with $\beta=0$ (b). }\label{fig10}
\end{figure}

Simple toy models allow one to test key ingredients for a new phenomenon that is the case here. 
In order to demonstrate the importance of nonlinear coupling between two oscillators, we set 
$\gamma=0$ for model (\ref{complex-osc}) and $\beta=0$ for model (\ref{m-vdp}). 
In both cases, we were not able to find any trace of dancing synchronization within many trials. 
Figure \ref{fig10} are what were typically observed with only conventional synchronizations where Fig.~\ref{fig10}(a) 
is the result of model (\ref{complex-osc}) with $\gamma=0$ and Fig.~\ref{fig10}(b) for model (\ref{m-vdp}) with 
$\beta=0$ while all other parameters are the same as those for Fig.~\ref{fig9} that show dancing 
synchronizations. Our studies show the importance of nonlinear couplings for the dancing synchronization.
	
\begin{figure*}
\includegraphics[width=1.6\columnwidth]{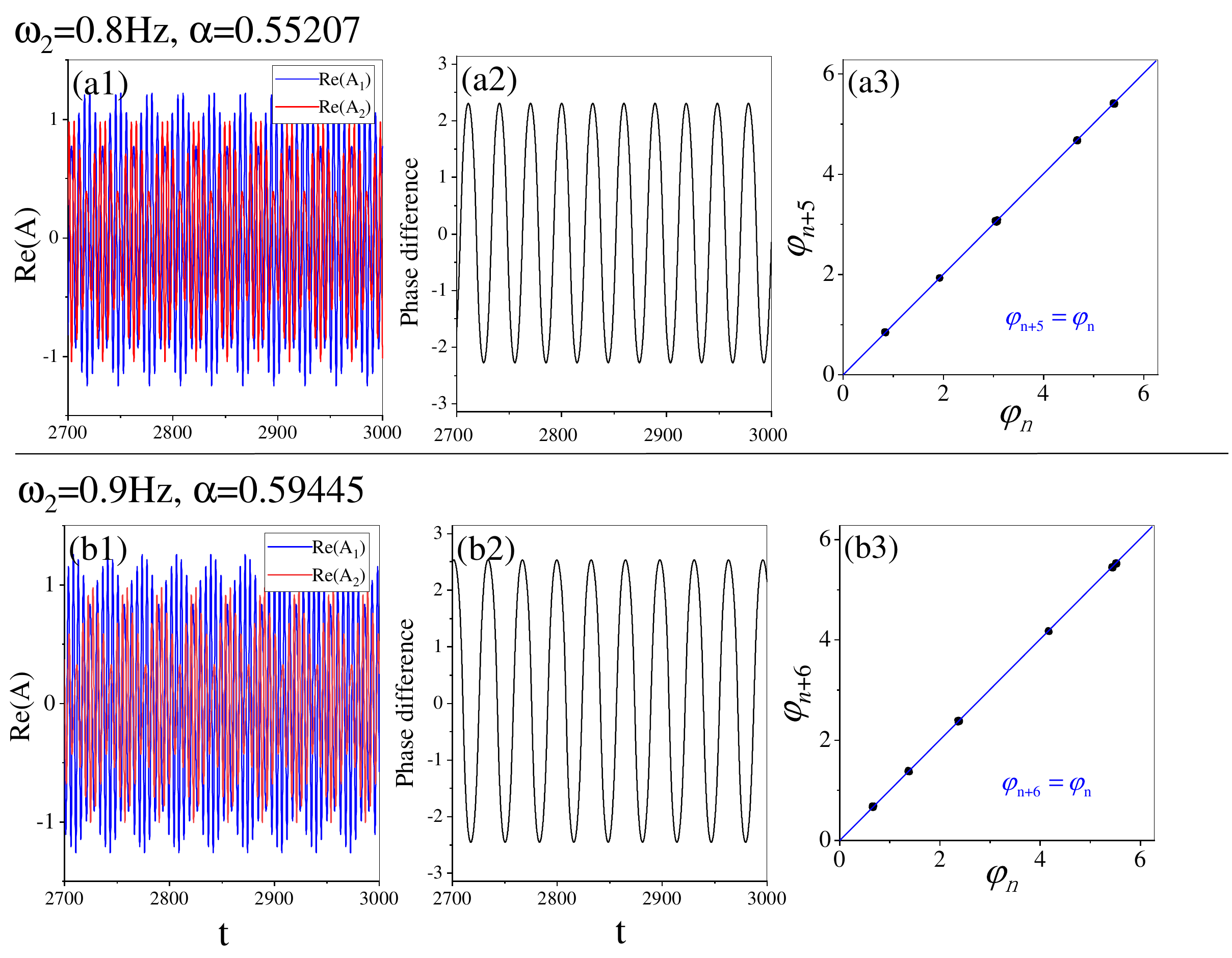}
\caption{Dancing synchronization of model (\ref{complex-osc}) with $\omega_2=0.8$ Hz, 
$\alpha=0.55207$ and $\omega_2=0.9\,$Hz, $\alpha=0.59445$, respectively. 
Panels from the left to the right are the time evolution of the two oscillators, 
the corresponding phase difference and the Poincaré map.}\label{fig11}
\end{figure*}

\begin{figure*}
\includegraphics[width=2.0\columnwidth]{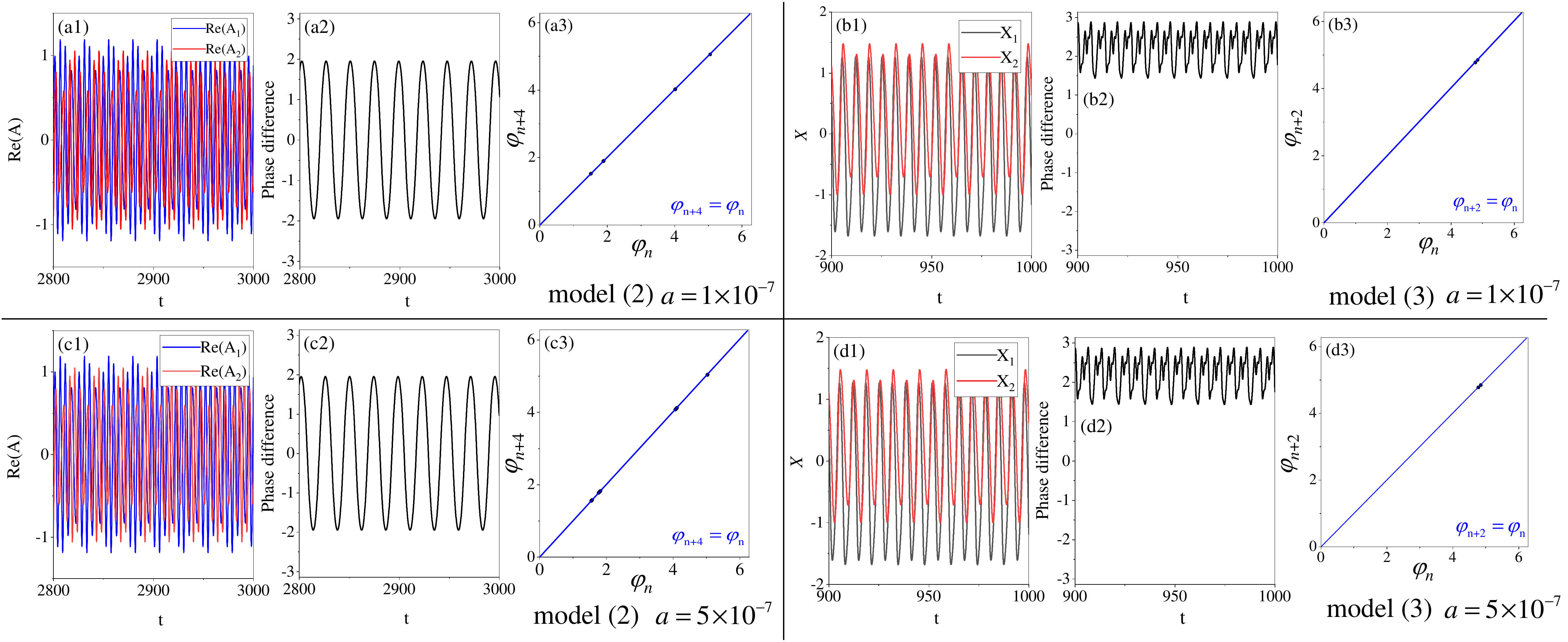}
\caption{Dancing synchronization of the complex amplitude model [model (\ref{complex-osc}) (a1-a3) 
for $a=1\times10^{-7}$ and (c1-c3) for $a=5\times10^{-7}$] and the Van der Pol model 
[model (\ref{m-vdp})] (b1-b3) for $a=1\times10^{-7}$ and (d1-d3) for $a=5\times10^{-7}$]. 
(a1, b1, c1, d1) are the time evolution of two oscillators. (a2, b2, c2, d2)
are the time evolution of phase difference. (a3, b3, c3, d3) are the Poincaré map. 
The periodical oscillation of phase difference and the the Poincaré map demonstrate the dancing 
synchronization under the noise.}\label{fig12}	
\end{figure*}

The dancing synchronization in the toy models is also robust against certain degree variation of 
parameters. As an example in model (\ref{complex-osc}) when the intrinsic frequency of the second 
oscillator and $\alpha$ change from 0.7 to 0.8 and from 0.59126 to 0.55207, respectively, dancing 
synchronization appears also as shown in Fig.~\ref{fig11}(a1$\sim$a3), in which the state returns 
to its starting point after moving around the origin of phase plane five turns Fig.~\ref{fig11}(a3). 
Similarly, if we change the intrinsic frequency of second oscillator from 0.7 to 0.9, and $\alpha$ 
from 0.59126 to 0.59445, dancing synchronization is still there as shown in Fig.~\ref{fig11}(b1$\sim$b3), 
in which the state returns to its starting point after six turns Fig.~\ref{fig11}(b3). 

A true natural phenomenon should be tolerable to thermal noise. To demonstrate that our dancing 
synchronization is insensitive to the thermal noise, we add a stochastic force to the original 
equations, e.g., the nonlinear dynamical equation of the complex amplitude model becomes
\begin{eqnarray}
\begin{split}
\dot{A_j}=&\lambda A_j\left(1-\left|A_j\right|\right)+i\left(\omega_jA_j+\alpha\left|A_j
\right|^2A_j\right)\\&+i\beta\sum_{k\neq j}^{n}\left({\widetilde{A}}_k-A_j\right)+
\gamma\sum_{k\neq j}^{n}{{\widetilde{A}}_k\left(1-|A_j|\right),}
	\end{split}
\end{eqnarray}
and the Van der Pol model becomes
\begin{equation}
	\begin{split}
		f_{ij}=\alpha(\widetilde{x}_j-x_i)+(j-i)\beta\sqrt{|x_i\widetilde{x}_j+\dot{x_i}\dot{x_j}-1|},
		\label{couple2}
	\end{split}
\end{equation}
where
${\widetilde{A}}_k=A_k+aS\left(t\right)$, ${\widetilde{x}}_j=x_j+aS\left(t\right)$ and $S(t)$ is a standard 
Gaussian stochastic process, $a$ measures the strength of random force. In simulations, an independent 
Gaussian-distributed random force of standard deviation $\sigma=1$ is assigned in each step ($\Delta t=4.7\times10^{-3}\,$s). 
We solved equations numerically with $a=1\times10^{-7}$ and $a=5\times10^{-7}$. The results of 
the complex amplitude model and the Van der Pol model are displayed in Fig.~\ref{fig12}. 
The Poincaré map (collecting data from 3000 periods) is slightly dispersed for both $a$'s. 
All return points fall around the line of $\varphi_{n+4}=\varphi_n$, which sustained our 
statement on the robustness of the dancing synchronizations. The dancing synchronization of the 
Van de Pol model is much resilient than that of complex amplitude model, as shown in Fig.~\ref{fig12} 
(b) and Fig.~\ref{fig12}(d) with $a=1\times10^{-7}$ and $a=5\times10^{-7}$ respectively.
\\
\par

\section{Conclusion}
\label{conclu}
In summary, a new type of synchronization, termed dancing synchronization, is 
observed in two STNOs coupled through spin waves and the static magnetic interaction. 
The two STNOs oscillate with the same period and their relative phase difference varies 
periodically with a common long period, different from all known 
synchronizations in which the relative phase of two nonlinear oscillators are fixed. 
We further demonstrated that the dancing synchronization is a general phenomenon that 
can also occur in the complex variable oscillation model used by Matheny and co-workers 
\cite{Mathenyeaav7932}, and in two coupled Van der Pol oscillators, as long as they are 
coupled reactively and dissipatively. The dancing synchronization exists in narrow 
parameter region between non-synchronization and in phase synchronization of two 
nonlinear oscillators.
	
\begin{acknowledgments}
This work is supported by the National Key Research and Development Program of China 
(Grant No. 2018YFB0407600 and 2016YFA0300702), the National Natural Science Foundation of China 
(Grant No. 12074301, 11774296, 11804266 and 11974296), the Key Research and Development 
Program of Shannxi (Grant No. 2019TSLGY08-04), Hong Kong RGC (Grants No. 16301518 and 16301619), 
and the Science Fund for Distinguished Young Scholars of Hunan Province (Grants No. 2018JJ1022).
\end{acknowledgments}

\bibliographystyle{apsrev4-1}
\bibliography{mibx}
	
\end{document}